\documentclass[10pt,conference]{IEEEtran}

\usepackage{cite}
\usepackage{amsmath,amssymb,amsfonts}
\usepackage{algorithm}
\usepackage{algpseudocode}
\usepackage{graphicx}
\usepackage{textcomp}
\usepackage{xcolor}
\usepackage{bbm} 
\usepackage{hyperref}
\usepackage{comment}
\usepackage{braket}
\usepackage{tikz}
\usepackage{quantikz}
\usepackage{cleveref}

\usepackage{graphicx}
\usepackage{subcaption}
\def\BibTeX{{\rm B\kern-.05em{\sc i\kern-.025em b}\kern-.08em
    T\kern-.1667em\lower.7ex\hbox{E}\kern-.125emX}}

\AtBeginDocument{%
\setlength{\textfloatsep}{4pt plus 1pt minus 2pt}%
\setlength{\floatsep}{4pt plus 1pt minus 1pt} 
\setlength{\intextsep}{6pt plus 1pt minus 2pt}%

\allowdisplaybreaks[4]

\setlength{\abovedisplayskip}{2pt plus 1pt minus 1pt}%
\setlength{\belowdisplayskip}{2pt plus 1pt minus 1pt}%
\setlength{\abovedisplayshortskip}{1pt plus 1pt minus 1pt}%
\setlength{\belowdisplayshortskip}{1pt plus 1pt minus 1pt}%
\setlength{\jot}{0.25pt}%
}
    
\newcommand{\inner}[2]{\langle {#1} \vert {#2} \rangle} 
\newcommand{\EXP}[1]{\left\langle {#1} \right\rangle} 
\newcommand{\abs}[1]{\left| {#1} \right|} 

\newcommand{\mysize}{{}} 

\newcommand{\corr}[1]{\textcolor{blue}{#1}} 
\newcommand{\comm}[1]{\textcolor{red}{#1}} 

\DeclareMathOperator{\re}{Re}
\DeclareMathOperator{\im}{Im}

\newcommand{\secref}[1]{Sec.~\ref{#1}}

\newcommand{\eqnref}[1]{(\ref{#1})}
\newcommand{\figref}[1]{Fig.~\ref{#1}}

\begin{document}

\title{Adiabatic Quantum Simulation of the Topological Su--Schrieffer--Heeger--Hubbard Model\\

\thanks{Special thanks to Pika Wang for providing access to NVIDIA computational resources, and to NVAITC and Taipei-1 for their invaluable support.}
}

\author{
\IEEEauthorblockN{
Ssu-Yi Chen\IEEEauthorrefmark{1}\IEEEauthorrefmark{3}\IEEEauthorrefmark{4}\IEEEauthorrefmark{5},
Bo-Hung Chen\IEEEauthorrefmark{1}\IEEEauthorrefmark{3}\IEEEauthorrefmark{4}\IEEEauthorrefmark{6},
Dah-Wei Chiou\IEEEauthorrefmark{1}\IEEEauthorrefmark{3}\IEEEauthorrefmark{4}\IEEEauthorrefmark{7},
and Jie-Hong Roland Jiang\IEEEauthorrefmark{1}\IEEEauthorrefmark{2}\IEEEauthorrefmark{3}\IEEEauthorrefmark{4}\IEEEauthorrefmark{8}
}
\IEEEauthorblockA{\IEEEauthorrefmark{1}Graduate Institute of Electronics Engineering,
National Taiwan University, Taipei, Taiwan}
\IEEEauthorblockA{\IEEEauthorrefmark{2}
Department of Electrical Engineering, National Taiwan University, Taipei, Taiwan}
\IEEEauthorblockA{\IEEEauthorrefmark{3}
Center for Quantum Science and Engineering, National Taiwan University, Taipei, Taiwan}
\IEEEauthorblockA{\IEEEauthorrefmark{4}
Physics Division, National Center for Theoretical Sciences, Taipei, Taiwan}
\IEEEauthorblockA{
Email: 
\IEEEauthorrefmark{5}r12943161@ntu.edu.tw,
\IEEEauthorrefmark{6}kenny81778189@gmail.com, 
\IEEEauthorrefmark{7}dwchiou@gmail.com, 
\IEEEauthorrefmark{8}jhjiang@ntu.edu.tw}
}

\maketitle

\begin{abstract}
We develop an adiabatic quantum simulation framework on gate-based quantum computers to probe topological signatures of the one-dimensional fermionic Su--Schrieffer--Heeger--Hubbard (SSHH) model. We present explicit quantum-circuit constructions for initial-state preparation and time evolution, together with a practical measurement protocol and classical post-processing procedure for extracting the many-body Berry phase and the spatial profile of the sublattice polarization. Using classical simulations of the proposed circuits, we demonstrate---for the first time within a genuine many-body framework---that the topological characteristics of the SSH model remain robust against weak Hubbard interactions but eventually break down as the chiral-symmetry-breaking component of the interaction exceeds a threshold. The required qubit number, gate complexity, measurement shots, and classical pre- and post-processing costs all scale polynomially with system size. Our results provide a proof-of-concept framework for probing topological properties of interacting many-body systems via adiabatic quantum simulation on future large-scale quantum computers.
\end{abstract}


\section{Introduction}

Simulating interacting many-body systems is widely regarded as one of the most promising near-term applications of quantum computation in the noisy intermediate-scale quantum (NISQ) era~\cite{Preskill2018,Bharti2022}. Classical numerical methods often become intractable because the Hilbert space grows exponentially with system size~\cite{Schollwoeck2005,Troyer2005,Kempe2006}. Quantum computers, by contrast, provide a natural platform for representing, evolving, and measuring quantum states~\cite{feynman1982}, potentially enabling the study of interacting quantum systems beyond the reach of classical approaches~\cite{Lloyd1996}.

A representative NISQ algorithm for such tasks is the \emph{variational quantum eigensolver} (VQE)~\cite{Peruzzo_2014,Kandala_2017_HEVQE}, a hybrid quantum--classical method that estimates the ground-state energy of a target quantum system by iteratively optimizing the parameters of a parameterized quantum circuit~\cite{Cerezo_2021_VQA,Tilly_2022}. The method benefits from the variational principle: if the approximated state deviates from the exact target state by $O(\varepsilon)$, the resulting energy error is typically $O(\varepsilon^2)$~\cite{McClean_2016_VHQA}. However, an accurate energy does not necessarily imply an accurate wavefunction. Unless the \emph{ansatz} (i.e., VQE circuit pattern) incorporates the relevant physical structure, the optimized state may fail to capture global, symmetry-protected, or topological properties of the true ground state~\cite{VaqueroSabater_2024,Sun_2023,Yoo_2025}. Consequently, for problems concerned with broader physical characteristics rather than energy estimation alone, VQE may not reliably reproduce the desired state.

An alternative approach is \textit{adiabatic quantum simulation}, in which the target state is obtained by slowly deforming a known ground state of a simple Hamiltonian into that of the target system~\cite{Farhi_2001,Albash_2018_AQC}. Provided that the evolution is sufficiently slow and the energy gap to excited states remains finite, the final state remains close to the desired ground state.

Adiabatic dynamics can be implemented either on analog quantum annealers or on gate-based quantum computers. Analog quantum annealers, such as D-Wave systems~\cite{Johnson_2011_DWaveQA}, naturally realize annealing dynamics but are typically limited to transverse-field Ising Hamiltonians~\cite{Kadowaki_1998_QA,Hauke_2020}. This restriction limits their direct applicability to interacting fermionic models~\cite{levy2022towards}, whose qubit representations generally involve noncommuting and often nonlocal Pauli operators beyond the Ising form~\cite{Bravyi_2002_FermionicQC,Seeley_2012_JWBK}. Gate-based quantum computers, by contrast, can in principle simulate arbitrary quantum dynamics~\cite{qute.201900052,fauseweh2024quantum}. In particular, adiabatic evolution can be realized within this framework, and may even be implemented digitally through fault-tolerant quantum computation~\cite{Aharonov_2008_AQCEquivalence}.

In this work, we investigate the one-dimensional fermionic Su--Schrieffer--Heeger--Hubbard (SSHH) model~\cite{Kivelson_1982} and demonstrate that nontrivial topological properties of many-body systems can be probed through adiabatic quantum simulation on gate-based quantum computers.

The SSHH model extends the Su--Schrieffer--Heeger (SSH) model~\cite{PhysRevLett.42.1698}---a paradigmatic model for one-dimensional topological insulators---by including the Hubbard interaction that accounts for on-site electron--electron Coulomb repulsion~\cite{Hubbard_1963}. In the absence of many-body interactions, the SSH model admits a simple single-particle description based on its band structure. The inclusion of the Hubbard term introduces particle--particle interactions and turns the system into a genuine many-body problem. The fermionic SSHH model therefore provides a natural platform for demonstrating the advantages of adiabatic quantum simulation and for exploring its limitations.

The SSHH model is also of broad interest in condensed matter physics, as it can exhibit a variety of phenomena depending on the model parameters, including antiferromagnetic order, topological and semimetallic phases, and superconductor--Mott-insulator transitions~\cite{Feng_2022,Le_2020}.

The SSH model has been extensively studied, including generalizations with arbitrary long-range hopping~\cite{Chen_2020}. The effects of many-body interactions have also been explored from single-particle or approximate perspectives~\cite{Di_Salvo_2024}, and quantum simulations of SSH-type many-body systems have been demonstrated on gate-based quantum computers~\cite{Chang_2025}. These studies suggest that the topological signatures of the SSH model are robust against weak interactions. However, this robustness has not yet been demonstrated from a genuinely many-body perspective.

Here we address this question by studying the SSHH model using adiabatic quantum simulation on gate-based quantum computers. Starting from a many-body ground state of the SSH model, we adiabatically introduce the Hubbard interaction. The initial state can be systematically prepared in quantum circuits based on pre-solved single-particle states, and the adiabatic evolution can likewise be implemented through a sequence of quantum gates. By extracting the many-body Berry phase and the spatial profile of the sublattice polarization through a practical measurement protocol and classical post-processing, we demonstrate--within a fully many-body framework---that the topological characteristics of the SSH model remain robust against weak Hubbard interactions but eventually break down when the chiral-symmetry-breaking component of the interaction becomes sufficiently strong.

Both the gate count and circuit complexity scale polynomially with system size for the initial-state preparation and the adiabatic evolution. The number of measurement shots required by the protocol and the runtime of the post-processing procedure also scale polynomially. Although our results are obtained from classical simulations for small systems, they provide a proof of concept that the same framework can be implemented on future large-scale quantum computers to investigate topological properties of larger interacting many-body systems.

\section{Adiabatic quantum simulation framework}\label{sec:adiabatic}
We consider a many-body quantum system governed by the Hamiltonian \(H = H_0 + H_1,\)
where \(H_0\) represents the one-body contribution and \(H_1\) the many-body interactions. The spectrum or ground state of \(H_0\), or at least the corresponding properties, can usually be obtained efficiently by classical methods and, in some cases, even analytically. Once \(H_1\) is included, however, determining the properties of the full Hamiltonian \(H\) rapidly becomes intractable beyond modest system sizes, because the number of degrees of freedom of many-body systems grows exponentially with the system size.

Adiabatic quantum simulation offers a promising approach to overcome this difficulty in accessing ground-state properties. The central idea is to vary the Hamiltonian slowly from \(H_0\) to \(H\), thereby transforming the ground state of \(H_0\) into that of \(H\). This can be implemented by initializing the system in the ground state of \(H_0\) and evolving it under a time-dependent Hamiltonian, for example,
\begin{equation}\label{Ht}
H(t) = H_0 + \frac{t}{T} H_1,
\qquad t \in [0,T],
\end{equation}
which continuously interpolates between \(H_0\) at \(t=0\) and \(H\) at \(t=T\). If the interpolation is sufficiently slow (i.e., if \(T\) is large enough), the adiabatic theorem ensures that the evolving state remains close to the instantaneous ground state of \(H(t)\) throughout the evolution. Consequently, the final state at \(t=T\) approximates the ground state of \(H\), allowing physical observables of the interacting many-body system to be extracted from measurements on the final state.


Let \(U(t,t')\) denote the time-evolution operator from \(t'\) to \(t\) generated by \(H(t)\). It satisfies the Schr\"odinger equation
\begin{equation}
i\hbar \frac{d}{dt}U(t,t') = H(t)U(t,t'),
\end{equation}
with the initial condition \(U(t',t')=\mathbbm{1}\). Formally, the solution can be written as the time-ordered (\(\mathcal{T}\)) exponential
\begin{equation}
U(T,0)=\mathcal{T}\exp\!\left(-\frac{i}{\hbar}\int_{0}^{T} H(t)\,dt\right).
\end{equation}

Although this time-ordered exponential is generally difficult to evaluate exactly, a useful approximation can be obtained by considering the infinitesimal propagator over a short interval \([t,t+\delta t]\). To first order in \(\delta t\), one finds
{\mysize
\begin{align}
&\quad U(t+\delta t,t)
= \mathbbm{1} - \frac{i}{\hbar}\int_t^{t+\delta t} H(t')\,dt' + O(\delta t^2) \\
&= \mathbbm{1} - \frac{i\,\delta t}{\hbar}
\left[H_0 + \frac{t'}{T} H_1\right]_{t'=t+\delta t/2} + O(\delta t^2) \nonumber\\
&=
\exp\!\left[-\frac{i\delta t}{\hbar} \left( \frac{t+\delta t/2}{T}\right) H_1\right]
\exp\!\left[-\frac{i\delta t}{\hbar} H_0\right]+ O(\delta t^2). \nonumber
\end{align}
}\par\noindent
This corresponds to the first-order Trotter decomposition, in which the Hamiltonian is treated as constant within each short interval and evaluated at the midpoint.

The full propagator \(U(T,0)\) can therefore be approximated by a product of many short-time propagators. Setting
\begin{equation}
\delta t = T/L,
\end{equation}
we obtain
\begin{align}
U(T,0)
&=
U(T, T -\delta t)\cdots U(\delta t, 0) + O(L\delta t^2)\notag\\
&\equiv U_L\, U_{L-1}\cdots U_{1} + O(T^2/L),
\end{align}
where the propagator for the \(\ell\)-th interval is
\begin{equation}\label{Uj}
U_\ell
=
\exp\!\left(-\frac{i\delta t}{\hbar} \frac{2\ell-1}{2L} H_1\right)
\exp\!\left(-\frac{i\delta t}{\hbar} H_0 \right).
\end{equation}

The requirement that the interpolation be sufficiently slow is quantified by the adiabatic condition, which demands that transitions between different instantaneous eigenstates of \(H(t)\) remain strongly suppressed throughout the evolution. Let \(\ket{n(t)}\) denote the instantaneous eigenstates of \(H(t)\) with corresponding eigenvalues \(E_n(t)\). Using time-dependent perturbation theory to estimate transition amplitudes yields the adiabatic condition
\begin{equation}\label{adiabatic condition}
\max_{t\in [0,T]}\max_{m\neq n}
\frac{\left|\bra{m(t)}\dot{H}(t)\ket{n(t)}\right|}{\left|E_m(t)-E_n(t)\right|^2}
\ll \frac{1}{\hbar}.
\end{equation}

For the interpolation in~\eqref{Ht}, the Hamiltonian varies at a constant rate, \(\dot{H}(t)=H_1/T.\)
The requirement that \(T\) be sufficiently large can therefore be expressed by the worst-case estimate \(T \gg\hbar \|H_1\|/\Delta_{\min}^2\),
where \(\|\cdot\|\) denotes the operator norm and \(\Delta_{\min}=\min_{t\in[0,T]}\min_{m\neq n}|E_m(t)-E_n(t)|\)
is the minimum instantaneous energy gap encountered along the interpolation path.

In practical applications this bound can often be tightened because the initial state is prepared in a specific form. For instance, if the system is initialized in the ground state, the relevant index pairs \((m,n)\) include only those with \(n\) corresponding to the ground state, which substantially reduces the bound.

It should be noted that this nondegenerate adiabatic analysis breaks down if the relevant instantaneous energy gap closes at some point during the evolution, for example because of an energy level crossing or a degeneracy enforced by symmetry. In such cases the standard adiabatic bound becomes singular, and the evolution must instead be analyzed within a degenerate adiabatic framework or along a modified interpolation path that maintains a finite gap. In many physical systems, however, the closing of the gap signals a genuine quantum phase transition. In these situations the singularity cannot be removed by such reformulations, as it reflects an intrinsic physical property of the system rather than a limitation of the analysis. This issue will be examined more concretely in the context of the SSHH model discussed below.

\section{Fermionic SSHH model}\label{sec:SSHH}

The Hamiltonian of the fermionic SSHH model can be expressed as
\begin{equation}
\label{H sshh}
H = H_\mathrm{SSH,\uparrow} + H_\mathrm{SSH,\downarrow} + H_\mathrm{Hubbard},
\end{equation}
where \(H_\mathrm{SSH,\uparrow}\) and \(H_\mathrm{SSH,\downarrow}\) describe the SSH model for the two spin sectors (\(\uparrow\) and \(\downarrow\)), and \(H_\mathrm{Hubbard}\) represents the on-site Hubbard interaction.



Consider a system of electrons in a one-dimensional lattice with \(N\) unit cells, each consisting of two sublattice sites, denoted \(A\) and \(B\). The lattice is assumed to satisfy either periodic boundary conditions (PBC) or open boundary conditions (OBC).

Let \(a_{js}^\dagger\) (\(a_{js}\)) and \(b_{js}^\dagger\) (\(b_{js}\)) denote the fermionic creation (annihilation) operators on the \(A\) and \(B\) sublattices, respectively, in the \(j\)-th unit cell, where \(s \in \{\uparrow,\downarrow\}\) labels the spin. These operators satisfy the canonical anticommutation relations
\begin{align}\label{anticommutation}
& \{a_{is}, a_{js'}\} = \{a_{is}^\dagger, a_{js'}^\dagger\}
 = \{b_{is}, b_{js'}\} = \{b_{is}^\dagger, b_{js'}^\dagger\} = 0, \notag\\
& \{a_{is}, a_{js'}^\dagger\} = \{b_{is}, b_{js'}^\dagger\}
 = \delta_{ij}\delta_{ss'}, 
 \qquad 
 \{a_{is}, b_{js'}^\dagger\} = 0,
\end{align}
which encode the Fermi--Dirac statistics of electrons. The corresponding number operators are
\begin{equation}
n_{Ajs} = a_{js}^\dagger a_{js}, \qquad
n_{Bjs} = b_{js}^\dagger b_{js}.
\end{equation}

The real-space Hamiltonian of the spinful SSH model is
\begin{align}
\label{H ssh}
H_{\mathrm{SSH},s}
=&
\sum_{j=1}^{N}
v \,b_{js}^\dagger a_{js}
+v^* \,a_{js}^\dagger b_{js}
\nonumber\\
&+
\sum_{j=1}^{N-1}
w \,b_{js}^\dagger a_{j+1,s}
+w^* \,a_{j+1,s}^\dagger b_{js}
\nonumber\\
&+  \text{periodic-boundary terms},
\end{align}
where \(v\) and \(w\) denote the intracell and intercell hopping amplitudes, respectively. The periodic-boundary contribution is 
\(w\, b_{Ns}^\dagger a_{1s}+w^*\, a_{1s}^\dagger b_{Ns}\)
for PBC, accounting for the hopping between the rightmost and leftmost cells, and \(0\) for OBC.
Equation~\eqref{H ssh} is invariant under the exchange of \(A\) and \(B\) sublattices, a property known as \emph{chiral symmetry}, which plays a crucial role in protecting the topological features of the model.

Since the spinful SSH model contains no many-body interactions, it admits a single-particle description. Its phases are characterized by the topological invariant \(\mathbbm{w}\), known as the \emph{winding number}. The system is a trivial insulator with \(\mathbbm{w}=0\) when \(|v|>|w|\), and a topological insulator with \(\mathbbm{w}=1\) when \(|v|<|w|\). At the critical point \(|v|=|w|\), the band gap closes and the system becomes metallic; this point marks the topological phase transition, where the winding number is ill-defined.

The Hubbard interaction is described by the real-space Hamiltonian
\begin{equation}
\label{H hubbard}
H_{\mathrm{Hubbard}}
=
\sum_{j=1}^{N}
\left(
U_A \,n_{Aj\uparrow}n_{Aj\downarrow}
+
U_B \, n_{Bj\uparrow}n_{Bj\downarrow}
\right),
\end{equation}
where \(U_A\ge0\) and \(U_B\ge0\) represent the on-site Coulomb repulsion on the \(A\) and \(B\) sublattices, respectively. In the conventional SSHH model one typically assumes \(U_A=U_B\equiv U\). Here we also consider the more general unbalanced case \(U_A\neq U_B\), which breaks the chiral symmetry and allows us to investigate how Hubbard interactions modify the topological features of the original SSH system. In the special case \(U_A = U_B \equiv U\) and \(v=w\), the SSHH model reduces to the standard one-dimensional Hubbard model.

In the strong-coupling limit \(U\gg |v|,|w|\), charge fluctuations are strongly suppressed and the system is effectively described by localized magnetic moments. When \(U/|v|\) is not large, the competition between hopping and Coulomb repulsion leads to nontrivial correlation effects.

In this work we focus on the transition between the trivial and topological insulating phases of the SSHH model. In particular, we investigate how robust the topological signatures of the original SSH model remain in the presence of particle--particle interactions, especially the unbalanced interactions that explicitly break chiral symmetry. This question is of interest because topological properties are generally expected to be robust against weak perturbations, yet their stability against many-body interactions has not been fully established.

\subsection{Adiabatic evolution of the SSHH system}\label{sec:adiabatic of SSHH}

Treating \(H_{\mathrm{SSH},\uparrow}+H_{\mathrm{SSH},\downarrow}\) as \(H_0\) and \(H_{\mathrm{Hubbard}}\) as \(H_1\) in~\eqref{Ht}, we can implement adiabatic quantum simulation on a quantum circuit. By repeatedly executing the circuit and measuring the final state, one can extract topological information about the SSHH model.

Note that \(H(t)\) does not induce spin flips. More precisely,
\begin{equation}\label{H spin commute}
[H,\hat{n}_s]
=[H_{\mathrm{SSH},s'},\hat{n}_s]
=[H_{\mathrm{Hubbard}},\hat{n}_s]=0,
\end{equation}
where the spin number operators are \(\hat{n}_s \equiv \sum_{\alpha\in\{A,B\}}\sum_j n_{\alpha js}\), 
with eigenvalues denoted by \(n_\uparrow\) and \(n_\downarrow\).
Consequently, the Hilbert space of the many-body system decomposes into subspaces with fixed spin populations, corresponding to the simultaneous eigenspaces of \(\hat{n}_\uparrow\) and \(\hat{n}_\downarrow\). These subspaces are invariant under \(H(t)\), and therefore the spin populations remain constant throughout the evolution governed by~\eqref{Ht}.

In realistic physical systems, however, spin-flip interactions are rarely completely absent. When the Hubbard interaction becomes sufficiently strong, the system may lower its energy by redistributing the spin populations, since spin flips can reduce on-site repulsion. From a mathematical perspective, each fixed-spin-population subspace labeled by \((n_{\uparrow},n_{\downarrow})\) therefore forms a dynamically disconnected sector of \(H(t)\). Different sectors, however, need not be separated by a nonzero energy gap.

This leads to an important caveat: the evolution described by~\eqref{Ht} may represent an adiabatic process only in an artificial sense, namely when the dynamics are restricted to a fixed \((n_{\uparrow},n_{\downarrow})\) sector. In general, starting from the ground state of \(H_0\), the final state obtained through~\eqref{Ht} need not coincide with the true ground state of the full Hamiltonian \(H(T)\).

Fortunately, in the situations considered in this paper the relevant ground states correspond to configurations in which the lower energy levels are fully occupied and separated from higher levels by a finite excitation gap. Because the occupied lower levels are completely filled, there is no available phase space for spin flipping within them. Consequently, provided that the Hubbard interaction remains sufficiently weak relative to the band gap and that \(T\) is sufficiently large, the evolution governed by~\eqref{Ht} represents a physically meaningful adiabatic process, and the final state faithfully approximates the ground state of the target Hamiltonian.

The initial state is taken to be the ground state of the spinful SSH model, obtained by filling the single-particle energy eigenstates in ascending order of energy. The corresponding many-body wavefunction is the Slater determinant of the occupied eigenstates, which equivalently can be written as
\begin{equation}\label{psi ground}
\ket{\Psi_{n_{\uparrow},n_{\downarrow}}}
=
a^\dag_{\epsilon_{1,\uparrow}}\dots a^\dag_{\epsilon_{n_{\uparrow},\uparrow}}
a^\dag_{\epsilon_{1,\downarrow}}\dots a^\dag_{\epsilon_{n_{\downarrow},\downarrow}}
\ket{0},
\end{equation}
where \(a^\dag_{\epsilon_{j,s}}\) is the creation operator for the \(j\)-th lowest single-particle eigenstate of \(H_{\mathrm{SSH},s}\), and \(n=n_{\uparrow}+n_{\downarrow}\) is the total number of electrons.

Because of~\eqref{H spin commute}, the spectrum of \(H(t)\) is invariant under exchanging \(\uparrow\) and \(\downarrow\), resulting in a two-fold degeneracy whenever \(n_\uparrow\neq n_\downarrow\). The lattice with \(N\) unit cells has \(4N\) single-particle degrees of freedom, corresponding to spin-up and spin-down electrons on the \(A\) and \(B\) sublattice sites of each cell.

Under PBC, the SSH model exhibits the spectrum of an insulator: the lower band, consisting of \(2N\) energy levels below \(0\), is separated from the upper band of \(2N\) levels above \(0\) by a nonzero excitation gap. At half filling (\(n=2N\)), the lower band is completely occupied, and the total many-body energy is separated from higher-energy states by a finite gap. Topological signatures can then be extracted from the many-body Berry phase of the \(n=2N\) system.

Under OBC, the SSH model with \(\mathbbm{w}=0\) exhibits a similar band structure: the lower band of \(2N\) levels below \(0\) remains separated from the upper band by a nonzero gap, so the \(n=2N\) state is again gapped from above.

In contrast, under OBC with \(\mathbbm{w}=1\), the SSH model exhibits \emph{edge modes} arising from its nontrivial topology. The bulk spectrum consists of \(2N-2\) levels below \(0\) and \(2N-2\) above \(0\), separated by a nonzero gap, together with four states near zero energy whose wavefunctions are localized at the boundaries (edges).

For the \(\mathbbm{w}=1\) case under OBC, if we consider the \(n=2N+2\) state, both the lower band and the four edge modes are fully occupied. The total energy of the system is then separated from higher-energy states by a finite gap, and topological signatures can be extracted from the edge localization of the sublattice polarization as the edge modes are all occupied.

For both the \(n=2N\) and \(n=2N+2\) states considered above, we have \(n_\uparrow=n_\downarrow\), so the spin-exchange degeneracy is lifted. Moreover, the total many-body energy is well separated from higher-energy states. Starting from these states, we can therefore safely perform the adiabatic simulation governed by~\eqref{Ht}; the resulting final state faithfully approximates the ground state of the target Hamiltonian, from which the topological signatures can be extracted.

\subsection{Topological characterization of the SSHH model}

Topological properties of the SSHH model can be characterized from several many-body perspectives, including the many-body Berry phase, the charge center, bulk charge polarization, and edge-state localization. These features are expected to be robust against weak perturbations, whether arising from one-body or many-body effects. Consequently, the SSHH model should retain the topological characteristics of the SSH model provided that the Hubbard interaction remains sufficiently weak.

For a many-body state $\ket{\Psi}$, the total particle number is
\begin{equation}
n := \bra{\Psi}\hat{n}\ket{\Psi}
\equiv
\bra{\Psi}(\hat{n}_\uparrow+\hat{n}_\downarrow)\ket{\Psi}.
\end{equation}
If $\ket{\Psi}$ belongs to an eigenspace of $\hat{n}$, it describes an $n$-particle system. 
For a lattice with $N$ unit cells, the filling ratio is defined as $n/N$. 
In our setting $n$ is an integer and $0\le n/N\le 4$. Writing the filling ratio in lowest terms as $\tilde n/\tilde N$ (with $\tilde n$ and $\tilde N$ coprime), we define the position operator
\begin{subequations}\label{X}
\begin{align}
\hat{X}
&:=\sum_{j=1}^{N}\sum_{s\in\{\uparrow,\downarrow\}}
j\left(a^\dagger_{j,s}a_{j,s}+b^\dagger_{j,s}b_{j,s}\right),\\
&\equiv
\sum_{j=1}^{N} j\left(
n_{Aj\uparrow}+n_{Aj\downarrow}+n_{Bj\uparrow}+n_{Bj\downarrow}
\right),
\end{align}
\end{subequations}
and the quantity
\begin{equation}\label{z}
z_N[\tilde{n}/\tilde{N}]
=
\bra{\Psi}
\exp\!\Big(\frac{i2\pi\tilde{N}}{N}\hat{X}\Big)
\ket{\Psi},
\end{equation}
which depends only on the filling ratio in lowest terms~\cite{Aligia_1999,Resta_1999}.

From $z_N[\tilde{n}/\tilde{N}]$ one defines the many-body (more precisely, $n$-body) Berry phase~\cite{Resta_1998,Aligia_1999,Resta_1999}
\begin{equation}\label{gamma}
\gamma_N = \im \ln z_N[\tilde{n}/\tilde{N}],
\end{equation}
which reduces to the Zak phase when $n=1$, i.e., the Berry phase obtained from integrating the Berry connection over the Brillouin zone of a single-particle band~\cite{PhysRevLett.62.2747}.

Under PBC, the expectation value of the position operator is related to $z_N$ by
\begin{equation}
\langle X_c\rangle
=
\frac{N}{2\pi\tilde N}\,
\im\ln z_N[\tilde n/\tilde N],
\end{equation}
which is commonly referred to as the charge center~\cite{Resta_1998,Aligia_1999,Resta_1999}.

For the single-particle SSH model, the Zak phase is related to the winding number via $\gamma_N=\mathbbm{w}\pi$~\cite{Asb_th_2016}, and therefore serves as an indicator of nontrivial topology. 
For a general $n$-electron state occupying different energy levels of the same single-particle SSH model, however, the $n$-body Berry phase defined above is not necessarily quantized. 
Nevertheless, when $\ket{\Psi}$ takes the form of~\eqref{psi ground} and corresponds to the half-filled ground state occupying all negative-energy levels, the $n$-body Berry phase again becomes quantized as $\gamma_N=\mathbbm{w}\pi$. 
This occurs because the occupied states completely fill the Brillouin zone of the negative-energy band.

In the thermodynamic limit $n,N\to\infty$ with fixed filling ratio, the many-body Berry phase is directly related to the charge polarization (dipole moment per unit cell), \(P_\mathrm{e}
=
e\langle X_c\rangle
=
\frac{e}{2\pi}\gamma_N .\)
The many-body Berry phase therefore corresponds to a physical observable and is, in principle, experimentally measurable.

Another hallmark of topological systems is the presence of robust edge states under OBC. 
In the single-particle SSH model, bulk–boundary correspondence implies that the trivial phase ($\mathbbm{w}=0$) hosts no edge states, whereas the topological phase ($\mathbbm{w}=1$) supports $2\mathbbm{w}$ zero-energy edge modes~\cite{Chen_2020}.

Edge modes can also be characterized from a many-body perspective. 
Under OBC we define the edge-occupation operator
\begin{equation}
\hat n_\mathrm{edge}
=
\sum_{\alpha\in\{A,B\}}
\sum_{s\in\{\uparrow,\downarrow\}}
(\hat n_{\alpha 1 s}+\hat n_{\alpha N s}).
\end{equation}
If single-particle energy levels are filled sequentially as in~\eqref{psi ground}, the expectation value $\langle\hat n_\mathrm{edge}\rangle$ behaves as a staircase function of $n$ when edge modes are present: it varies smoothly with $n$ except at certain critical values where it jumps abruptly as an edge mode becomes occupied.

Since edge states themselves are not directly measurable, it is useful to consider experimentally accessible quantities. 
For this purpose we introduce the sublattice polarization operator
\begin{equation}
P^{\mathrm{e}}_j
=
\sum_{s\in\{\uparrow,\downarrow\}}
(n_{Ajs}-n_{Bjs}).
\end{equation}
Although $\langle P^{\mathrm{e}}_j\rangle$ vanishes for the half-filled ground state ($n=2N$), it becomes nonzero at the boundary cells for the ground state with two additional electrons ($n=2N+2$) if edge modes are present. 
This occurs because the edge states are fully occupied in this configuration. 
The spatial profile of $\langle P^{\mathrm{e}}_j\rangle$ under OBC therefore provides an experimentally meaningful indicator of the system's topological character.

These topological signatures of the SSH model are expected to persist under weak perturbations, including the Hubbard interaction introduced in the SSHH model. 
When the interaction strength is sufficiently small compared with the hopping amplitudes, the single-particle description remains a good approximation. 
In this regime the Hubbard term can be treated within a mean-field approximation. 
For the state \eqref{psi ground}, the effective Hubbard Hamiltonian becomes
{\mysize
\begin{align}\label{mean field approx}
&\quad H_\mathrm{Hubbard}^{\mathrm{eff}} \\
&= \sum_{j=1}^{N} \bigg[
U_A \left( 
\EXP{n}_{Aj\uparrow}n_{Aj\downarrow} 
+  n_{Aj\uparrow}\EXP{n}_{Aj\downarrow}
+ \EXP{n}_{Aj\uparrow}\EXP{n}_{Aj\downarrow}\right)\notag\\
&\quad+U_B\left( 
\EXP{n}_{Bj\uparrow}n_{Bj\downarrow} 
+  n_{Bj\uparrow}\EXP{n}_{Bj\downarrow}
+ \EXP{n}_{Bj\uparrow}\EXP{n}_{Bj\downarrow}\right).
\bigg]\notag
\end{align}
}\par\noindent
which effectively adds on-site one-particle potentials to $H_{\mathrm{SSH},\uparrow}$ and $H_{\mathrm{SSH},\downarrow}$.

Therefore, as long as the difference \(\Delta U := U_B-U_A\)
remains sufficiently small, the many-body Berry phase and the edge sublattice polarization inherited from the SSH model remain intact even though chiral symmetry is explicitly broken when $\Delta U\neq0$. 
By examining these indicators for the $n=2N$ and $n=2N+2$ ground states, we can demonstrate that the nontrivial topology persists under weak perturbations but eventually breaks down when interactions become sufficiently strong.

As discussed in \secref{sec:adiabatic of SSHH}, the adiabatic evolution governed by~\eqref{Ht} may fail to produce the true ground state of the target Hamiltonian because the dynamics are restricted to a fixed-spin-population sector $(n_\uparrow,n_\downarrow)$. 
To ensure that the evolution corresponds to a physically meaningful process, the Hubbard interaction must remain weak enough so that spin flips do not occur during the evolution.

For the $n=2N$ and $n=2N+2$ ground states, this requirement translates into a constraint on the energy cost of flipping a spin, characterized by $\max\{U_A,U_B,|\Delta U|\}$. The relevant energy scale is the single-particle band gap of the SSH model,
\begin{equation}
\Delta_{\mathrm{gap}}
=
2\min_k|v+e^{-ik}w|
=
2\min\{|v+w|,|v-w|\}.
\end{equation}
For the $n=2N$ ground state, the energy cost of a spin flip must be smaller than $\Delta_{\mathrm{gap}}$, whereas for the $n=2N+2$ ground state it must be smaller than $\Delta_{\mathrm{gap}}/2$. A sufficient condition is therefore estimated as
\begin{equation}\label{weak disturbance condition}
\max\{U_A,U_B,|\Delta U|\}
<
\min\{|v+w|,|v-w|\}.
\end{equation}

In \secref{sec:simulation}, we perform numerical simulations of the adiabatic evolution in this parameter regime and examine the resulting topological signatures of the SSHH model.\footnote{When $\Delta U=0$, evolving the half-filled ($n=2N$) ground state of $H_0$ under~\eqref{Ht} preserves the same topological characteristics in both the many-body Berry phase and the sublattice polarization profile regardless of the value of $U_A=U_B$, since the Hubbard interaction merely adds a uniform on-site potential. However, beyond the condition in~\eqref{weak disturbance condition}, the final state obtained from the adiabatic evolution can deviate considerably from the true ground state of the target Hamiltonian.}

\section{Quantum-circuit implementation of adiabatic simulation}\label{sec:implementation}
To implement adiabatic quantum simulation on a gate-based quantum computer, the continuous evolution generated by $H(t)$ must be approximated by a sequence of quantum gates. For the simulation to be efficient, the total qubit number and gate complexity should scale at most polynomially with the size of the simulated system. Likewise, the measurement protocol must remain scalable, requiring only polynomially many shots, and the classical post-processing used to estimate observable expectation values from measurement outcomes must also be computationally efficient. In this section, we develop such a framework for the SSHH model.

We begin by applying the Jordan--Wigner transformation to express the second-quantized fermionic Hamiltonian in terms of Pauli operators. Using the resulting qubit representation, we then construct explicit quantum circuits for preparing the initial state and implementing the time evolution generated by $H(t)$, both with polynomial gate depth. Finally, we describe a practical measurement protocol and post-processing procedure for extracting the many-body Berry phase and the spatial profile of the sublattice polarization.

\subsection{Jordan--Wigner transformation}

We employ the Jordan--Wigner transformation to express the fermionic algebra in terms of Pauli matrices. We denote the Pauli-$Z$ eigenstates as $\ket{Z+}\equiv\ket{0}$ and $\ket{Z-}\equiv\ket{1}$, corresponding to an empty and an occupied fermionic state, respectively.

More explicitly, for a system consisting of $N$ unit cells, the Jordan--Wigner transformation maps the fermionic creation and annihilation operators according to
{\mysize
\begin{subequations}\label{JW}
\begin{align}
a_{j\uparrow}   &\rightarrow \frac{1}{2} (X_{Aj\uparrow} + iY_{Aj\uparrow})               \prod_{k=1}^{j-1} Z_{Ak\uparrow} Z_{Bk\uparrow},\\
b_{j\uparrow}   &\rightarrow \frac{1}{2} (X_{Bj\uparrow} + iY_{Bj\uparrow}) Z_{Aj\uparrow}\prod_{k=1}^{j-1} Z_{Ak\uparrow} Z_{Bk\uparrow},\\
a_{j\downarrow} &\rightarrow \frac{1}{2} (X_{Aj\downarrow} + iY_{Aj\downarrow})                  P^z_\uparrow\prod_{k=1}^{j-1} Z_{Ak\downarrow} Z_{Bk\downarrow},\\
b_{j\downarrow} &\rightarrow \frac{1}{2} (X_{Bj\downarrow} + iY_{Bj\downarrow})P^z_\uparrow Z_{Aj\downarrow} \prod_{k=1}^{j-1} Z_{Ak\downarrow} Z_{Bk\downarrow},
\end{align}
\end{subequations}
}\par\noindent
where the product is understood to be the identity for $j=1$. The operator $P_s^z$ represent the fermionic parity of the entire spin-$s$ sector, defined as
\begin{equation}
P_s^z := \prod_{k=1}^{N} Z_{Aks} Z_{Bks}.
\end{equation}

Under this mapping, the local number operators take the simple diagonal form
\begin{subequations}\label{n JW}
\begin{align}
n_{Ajs} \equiv a_{js}^\dagger a_{js} &= \left(I_{Ajs} - Z_{Ajs}\right)/2,\\
n_{Bjs} \equiv b_{js}^\dagger b_{js} &= \left(I_{Bjs} - Z_{Bjs}\right)/2,
\end{align}
\end{subequations}
whereas the intracell, intercell, and boundary hopping terms are mapped as
{\mysize
\begin{subequations}\label{hopping JW}
\begin{align}
b_{js}^\dagger a_{js} 
&=(X_{Bjs} - iY_{Bjs})( X_{Ajs} +iY_{Ajs})/4,\\
b_{js}^\dagger a_{j+1s} 
&= (X_{Bjs} -iY_{Bjs}) (X_{Aj+1s} + iY_{Aj+1s})/4,\\
b_{Ns}^\dagger a_{1s} &=  (X_{BNs}-iY_{BNs}) (X_{A1s} + iY_{A1s}) P_s^z/4.
\end{align}
\end{subequations}
}

Substituting \eqnref{hopping JW} into \eqnref{H ssh}, we obtain the Pauli-operator representation of the spinful SSH Hamiltonian,
{\mysize
\begin{align}
\label{H ssh JW}
 H_{\mathrm{SSH},s }
&=
-\frac{1}{2}
\Bigl\{
\sum_{j=1}^{N}
\Bigl[
\operatorname{Re}(v)
\bigl(X_{Ajs}X_{Bjs}+Y_{Ajs}Y_{Bjs}\bigr)
\notag\\
&\qquad\qquad\qquad
+\operatorname{Im}(v)
\bigl(X_{Ajs}Y_{Bjs}-Y_{Ajs}X_{Bjs}\bigr)\Bigr]
\notag\\
&\quad
+\sum_{j=1}^{N-1}\Bigl[
\operatorname{Re}(w)
\bigl(
X_{Aj+1s}X_{Bjs}+Y_{Aj+1s}Y_{Bjs}
\bigr)
\notag\\
&\qquad\qquad
+\operatorname{Im}(w)
\bigl(X_{Aj+1s}Y_{Bjs}-Y_{Aj+1s}X_{Bjs}\bigr)
\Bigr]
\notag\\
&\quad + \text{periodic-boundary terms} \Bigr\},
\end{align}
}\par \noindent
where the periodic-boundary contribution is $0$ for OBC and for PBC given by
\begin{align}
&\Bigl[\operatorname{Re}(w)
\bigl(X_{A1s}X_{BNs}+Y_{A1s}Y_{BNs}\bigr)
\notag\\
&\qquad
+\operatorname{Im}(w)
\bigl(X_{A1s}Y_{BNs}-Y_{A1s}X_{BNs}\bigr) 
\Bigr] P_s^z,
\end{align}
accounting for the hopping between the 1st and $N$-th cells.

Similarly, substituting \eqref{n JW} into \eqref{H hubbard}, we obtain the Pauli-operator representation of the Hubbard Hamiltonian,
{\mysize
\begin{align}\label{Hhubbard JW}
H_\mathrm{Hubbard}  
&=
\frac{U}{4}\sum_{s}
\sum_{j=1}^{N}
\Bigl[
(I_{Aj\uparrow}-Z_{Aj\uparrow})(I_{Aj\downarrow}-Z_{Aj\downarrow})
\notag\\
&\qquad\qquad
+(I_{Bj\uparrow}-Z_{Bj\uparrow})(I_{Bj\downarrow}-Z_{Bj\downarrow})
\Bigr]. 
\end{align}
}

In the same manner, we consider the propagator on the $\ell$-th interval defined in \eqnref{Uj},
\begin{equation}\label{Uj SSHH}
U_\ell = U_{\ell}^{\mathrm{Hubbard}}\, U_{\ell}^{\mathrm{SSH},\uparrow}\, U_{\ell}^{\mathrm{SSH},\downarrow}.
\end{equation}
Substituting \eqnref{H ssh JW} into \eqnref{Uj SSHH} and applying the first-order Trotter decomposition, the spinful SSH propagator in the Pauli-operator representation is given by
{\mysize
\begin{align} \label{U ssh p}
&\quad U_{\ell}^{\mathrm{SSH},s}\\
&=            \exp\bigg(\frac{\delta t \re(w)}{-2i}                  (X_{A1s}   X_{BLs} + Y_{A1s}   Y_{BLs})P_s^z\bigg)\notag\\
&\quad \times \exp\bigg(\frac{\delta t \re(w)}{-2i} \sum_{i=1}^{N-1}(X_{Ai+1s} X_{Bis} + Y_{Ai+1s} Y_{Bis})\bigg)\notag\\
&\quad \times \exp\bigg(\frac{\delta t \re(v)}{-2i} \sum_{i=1}^{N}  (X_{Ais}   X_{Bis} + Y_{Ais}   Y_{Bis})\bigg)\notag\\
&\quad \times \exp\bigg(\frac{\delta t \im(w)}{-2i}                  (X_{A1s}   Y_{BLs} - Y_{A1s}   X_{BLs})P_s^z\bigg)\notag\\
&\quad \times \exp\bigg(\frac{\delta t \im(w)}{-2i} \sum_{i=1}^{N-1}(X_{Ai+1s} Y_{Bis} - Y_{Ai+1s} X_{Bis})\bigg)\notag\\
&\quad \times \exp\bigg(\frac{\delta t \im(v)}{-2i} \sum_{i=1}^{N}  (X_{Ais}   Y_{Bis} -Y_{Ais}   X_{Bis})\bigg) + O(\delta t^2) \notag
\end{align}
}\par \noindent
where the error term represents the approximation inaccuracy of the first-order Trotter decomposition.

Within a fixed $(n_\uparrow,n_\downarrow)$ sector, the whole system has a definite parity. Consequently, \(P^z_s\) acts as a c-number and can be replaced by \(P^z_s \rightarrow (-1)^{n_s}\). All the terms involving \(P^z_s\) appear only in PBC and are absent in OBC. 

Similarly, substituting \eqnref{Hhubbard JW} into \eqnref{Uj SSHH}, we have
{\mysize
\begin{align}
&\quad U_{\ell}^{\mathrm{Hubbard}}\notag\\
&=\exp\Bigl(
\frac{\delta t U_A}{4i}\left(\frac{2\ell-1}{2L}\right) \sum_{j=1}^{N}  
(I_{Aj\uparrow} - Z_{Aj\uparrow})(I_{Aj\downarrow} - Z_{Aj\downarrow})  \Bigr)\notag\\
&\,\times\exp\Bigl(\frac{\delta t U_B}{4i}\left(\frac{2\ell-1}{2L}\right)  \sum_{j=1}^{N} 
(I_{Bj\uparrow} - Z_{Bj\uparrow})(I_{Bj\downarrow} - Z_{Bj\downarrow})\Bigr) \notag\\
&\,+ O(\delta t^2), 
\end{align}
}\par \noindent
where the error term again arises from the first-order Trotter decomposition. 

Under this mapping, each fermionic mode $(\alpha,j,s)$ is associated with a qubit whose computational basis encodes its occupation number. Consequently, the many-body wavefunction on a lattice with $N$ unit cells can be represented using $4N$ qubits, enumerated as \(q\in\{0,\ldots,4N-1\}\). The presence (absence) of a fermion in the mode $(\alpha,j,s)$ is mapped to the qubit state $\ket{1}$ ($\ket{0}$) of the corresponding qubit, for example, according to the explicit prescription
\begin{subequations}\label{qubit mapping}
\begin{align}
q(A,j,\uparrow )  &  = 2j-2,&
q(A,j,\downarrow) &  = 2N +2j-2, \\
q(B,j,\uparrow )  &  = 2j-1,&
q(B,j,\downarrow )&  = 2N +2j-1,
\end{align}    
\end{subequations}
where the $4N$ qubits are partitioned into two segments corresponding to the spin-up and spin-down sectors.

For later use, we define three types of two-qubit Pauli-rotation gates acting on the qubit pair \((i,j)\):
{\mysize
\begin{align}
R_{i,j}(\theta) 
& 
= \exp{\Bigl(\left( X_i X_j + Y_i Y_j\right)\theta/2i\Bigr)},\label{R}\\
G_{i,j}(\theta) 
&
= \exp{\Bigl(\left( X_i Y_j -Y_iX_j \right)\theta/2i\Bigr)},\label{G}\\
CP_{i,j}(\theta)
& 
= \exp{\Bigl(\left(I_i - Z_i\right)\left(I_j - Z_j\right)\theta/4i\Bigr)}.\label{CP}
\end{align}
}\par\noindent
In principle, each of these gates can be decomposed into two CNOT gates and at most four single-qubit rotation gates.

With these definitions, the gate-level propagator for the \(\ell\)-th time interval is implemented as
{\mysize
\begin{align}
&U_\ell
=  \prod_{j=0}^{N-1} 
CP_{2j, 2j+2N}(\phi_{A,\ell})\, 
CP_{2j+1, 2j+2N+1}(\phi_{B,\ell}) \notag\\
&\ \times \prod_{j=0,\,2N}
R_{j+2N-1,j}((-1)^{n_s}\theta_w)\,
G_{j+2N-1,j}((-1)^{n_s}\vartheta_w) \notag\\
&\quad \times\prod_{j=1}^{2N-2} 
R_{j,j+1}(\theta_w)\, 
R_{j+2N,j+2N+1}(\theta_w) \notag\\
&\quad \times\prod_{j=1}^{2N-2} 
G_{j,j+1}(\vartheta_w) \,
G_{j+2N,j+2N+1}(\vartheta_w)\notag\\
&\quad \times\prod_{j=0}^{4N-1} 
R_{j,j+1}(\theta_v)\, 
G_{j,j+1}(\vartheta_v), 
\label{U ell gate}
\end{align}
}\par \noindent
where
\begin{subequations}
\begin{align}
\theta_v &=-\delta t \re(v), & \vartheta_v&=-\delta t\im(v),\\
\theta_w &=-\delta t \re(w), & \vartheta_w&=-\delta t\im(w),\\
\phi_{A,\ell}&= \delta t U_A \left(\frac{2\ell-1}{2L}\right),&
\phi_{B,\ell}&= \delta t U_B\left(\frac{2\ell-1}{2L}\right).
\end{align}    
\end{subequations}

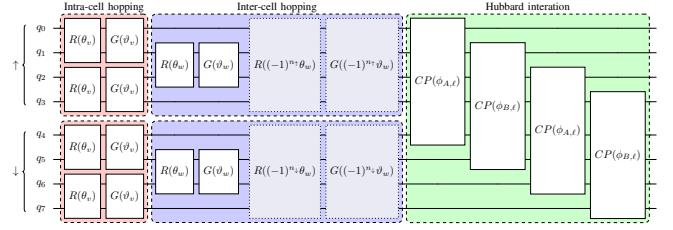
\begin{figure}[t]
\centering
\resizebox{0.48\textwidth}{!}{
\begin{quantikz}[row sep=4pt, column sep=5pt]
\lstick[4]{$\uparrow$}\midstick{$q_0$}&& \gate[2]{R(\theta_v)}  
\gategroup[4,steps=2, style={dashed, rounded corners, fill=red!20, inner sep= 0.2pt}, background]{Intra-cell hopping}
&\gate[2]{G(\vartheta_v)}
&[6pt] 
\gategroup[4,steps=4, style={dashed, rounded corners, fill=blue!20, inner sep= 0.2pt}, background]{Inter-cell hopping}
&
&[5pt] \gate[4,style={dotted, fill=gray!10, fill opacity=0.8}]{R((-1)^{n_\uparrow}\theta_w)}
&\gate[4,style={dotted, fill=gray!10, fill opacity=0.8}]{G((-1)^{n_\uparrow}\vartheta_w)}
&[6pt]\gate[5]{CP(\phi_{A,\ell})} \gategroup[8,steps=4, style={dashed, rounded corners, fill=green!20, inner sep= 0.2pt}, background]{Hubbard interation}
&&&&&  \\
\midstick{$q_1$}  &&&&\gate[2]{R(\theta_w)} & \gate[2]{G(\vartheta_w)} &  &  & & \gate[5]{CP(\phi_{B,\ell})} &  & & &  \\
\midstick{$q_{2}$}&& \gate[2]{R(\theta_v)} & \gate[2]{G(\vartheta_v)} &  &  & &  & & & \gate[5]{CP(\phi_{A,\ell})}&  &  &  \\
\midstick{$q_{3}$}&&&   & &  & & &    &  &  & \gate[5]{CP(\phi_{B,\ell})}&   &  \\[8pt] %
\lstick[4]{$\downarrow$}\midstick{$q_{4}$}&&
\gate[2]{R(\theta_v)}\gategroup[4,steps=2,style={dashed, rounded corners, fill=red!20, inner sep= 0.1pt},background]{} &
\gate[2]{G(\vartheta_v)} &  
\gategroup[4,steps=4, style={dashed, rounded corners, fill=blue!20, inner sep= 0.1pt}, background]{}&&
\gate[4,style={dotted, fill=gray!10, fill opacity=0.8}]{R((-1)^{n_\downarrow}\theta_w)}& 
\gate[4,style={dotted, fill=gray!10, fill opacity=0.8}]{G((-1)^{n_\downarrow}\vartheta_w)}&&&&&&  \\
\midstick{$q_{5}$}&&&&\gate[2]{R(\theta_w)}&\gate[2]{G(\vartheta_w)}&&&&&&&&  \\
\midstick{$q_{6}$}&&   \gate[2]{R(\theta_v)}& \gate[2]{G(\vartheta_v)}&&&&&&&& &  & \\
\midstick{$q_7$}&&&&&& &  &  &  & & &  &
\end{quantikz}
}
\caption{Quantum circuit implementation of a single Trotter step for the model with $N=2$ unit cells. Qubits $q_0$ to $q_3$ and $q_4$ to $q_7$ encode the spin-up and spin-down orbitals, respectively. The circuit is partitioned into three main stages: (1) The red shaded part represents the intra-cell hopping operations, utilizing $R$ and $G$ gates to realizing the real and imaginary components. (2) The blue shaded part represents the inter-cell hopping operations. The $R$ and $G$ gates in the gray boxes represent the periodic-boundary hopping terms, present only in PBC. (3) The green shaded part represents the Hubbard interaction, implemented via Controlled-Phase gates, which entangle the corresponding spin-up and spin-down orbitals at each site.} 
\label{fig:circ U}
\end{figure}

The circuit representation of \(U_\ell\) is shown in \figref{fig:circ U}. 
Under PBC, each propagator contains \(4N\) \(R_{ij}\) gates, \(4N\) \(G_{ij}\) gates, and \(2N\) \(CP_{ij}\) gates, giving a total of \(10N\) two-qubit Pauli-rotation gates. 
Under OBC, the corresponding counts are \(4N-2\), \(4N-2\), and \(2N\), respectively, giving a total of \(10N-4\) gates. 
Since each such gate decomposes into two CNOT gates and four single-qubit rotation gates, implementing \(L\) propagators requires \(2L(10N)\) CNOT gates and \(4L(10N)\) single-qubit rotation gates for PBC, and \(2L(10N-4)\) CNOT gates and \(4L(10N-4)\) single-qubit rotation gates for OBC. 
Therefore, the total gate count scales as \(O(NL)\).

\subsection{Initial state preparation}
To prepare the fermionic many-body ground state of the SSH model defined in \eqref{psi ground} in the lattice-site basis, we employ the state-preparation framework introduced in Ref.~\cite{Jiang_2018} and implemented in the Python library \textit{OpenFermion}~\cite{mcclean2019openfermionelectronicstructurepackage, OpenFermion}. 
Within this framework, the target state can be prepared by a quantum circuit whose depth scales polynomially  with the number of qubits. 
Below, we explain how this preparation routine is integrated into our circuit construction.

Suppose that the target state, obtained either analytically or numerically, takes the form given in \eqnref{psi ground}. 
Our goal is to prepare this state in the lattice-site basis defined by the creation operators \(a^\dagger_{js}\) and \(b^\dagger_{js}\) on quantum circuits. 
For this purpose, we express the single-particle creation operator \(a^\dagger_{\epsilon s}\) in terms of the lattice-site operators as
\begin{equation}
a^\dagger_{\epsilon_{\jmath,s} } = 
\sum_{j}
\left( \inner{A,j,s}{\epsilon_{\jmath,s}}\, a^\dagger_{js} + \inner{B, j,s}{\epsilon_{\jmath,s}}\, b^\dagger_{js}\right), 
\end{equation}
where \(\ket{\epsilon_s}\) denotes the single-particle energy eigenstate in the spin-\(s\) sector. 
Substituting this transformation into \eqnref{psi ground}, the many-body state becomes
{\mysize
\begin{align}\label{psi trans c2d}
\ket{\Psi_{n_\uparrow,n_\downarrow}} 
=\prod_{s\in\{\uparrow,\downarrow\}}
\prod_{\epsilon = \epsilon_1}^{\epsilon_{n_s}}
    \Bigl[\sum_{j}
        &\Bigl( \inner{A,j,s}{\epsilon_{\jmath,s}} \, a^\dagger_{js}\notag\\ 
        &\quad +\inner{B,j,s}{\epsilon_{\jmath,s}}\, b^\dagger_{js}\Bigr) \Bigr]
    \ket{0}.
\end{align}}
In the logical-qubit representation, the index \((\epsilon_{\jmath,s})\) appearing in \eqnref{psi ground} is mapped to the qubit index \(p\equiv p(\epsilon_{\jmath,s})\) as
\begin{equation}
p(\epsilon_{\jmath,\uparrow})=\jmath-1,\qquad 
p(\epsilon_{\jmath,\downarrow})=2N+\jmath-1,
\end{equation}
following the same indexing convention used by \textit{OpenFermion}.

Although a direct preparation of the state in \eqref{psi trans c2d} appears prohibitively complicated for large systems, the method of Ref.~\cite{Jiang_2018} provides an efficient construction by exploiting the fermionic anticommutation relations. 
The key observation is that the antisymmetric structure imposed by the Pauli exclusion principle allows the state-preparation problem to be reduced to a sequence of structured matrix decompositions. 
As a result, both the classical preprocessing cost and the quantum circuit construction depend only on the relevant rectangular transformation matrix
\begin{equation}
 Q_{pq}
 =
 \inner{\epsilon_{\jmath,s}}{\alpha,j,s},
 \qquad
 p\in S_p,\quad
 q\in S_q,
\end{equation}
where \(S_p\) indexes the occupied single-particle modes and \(S_q\) indexes the lattice-site modes used in the qubit representation. 
Accordingly, \(|S_p|\) is the number of fermions, whereas \(|S_q|\) is the number of lattice-site modes included in the mapping, which is typically the full single-particle Hilbert-space dimension of the system. 

This state-preparation procedure first applies \(n\) \(X\) gates to prepare the occupation pattern in \eqnref{psi ground}. 
It then implements the basis transformation from \eqnref{psi ground} to \eqnref{psi trans c2d} using \(|S_p|(|S_q|-|S_p|)\) \(G_{ij}\) gates. 
For a system with \(N_f\) single-particle degrees of freedom, the half-filled case \(|S_p|=N_f/2\) and \(|S_q|=N_f\) therefore requires \(N_f^2/4\) \(G_{ij}\) gates. 
Since each \(G_{ij}\) gate is decomposed into two CNOT gates and four single-qubit rotation gates, the corresponding gate counts are \(N_f^2/2\) CNOT gates and \(N_f^2\) single-qubit rotation gates, in addition to the \(N_f/2\) initial \(X\) gates. 
Equivalently, if the initial \(X\) gates are included among the single-qubit gates, the total number of single-qubit gates is \(N_f(N_f+1/2)\).

In the SSH model considered here, the two spin sectors are independent in the initial-state preparation. 
The transformations in the spin-up and spin-down sectors can therefore be implemented in parallel. 
Although the full system contains \(4N\) single-particle degrees of freedom, each spin sector contains only \(2N\) lattice-site modes. 
At half filling, \(|S_p|_{\uparrow}=n_\uparrow=N\), \(|S_p|_{\downarrow}=n_\downarrow=N\), and \(|S_q|_{\uparrow}=|S_q|_{\downarrow}=2N\). 
Thus, each spin sector requires \(N^2\) \(G_{ij}\) gates, corresponding to \(2N^2\) CNOT gates and \(4N^2\) single-qubit rotation gates, plus \(N\) initial \(X\) gates. 
Since the two spin sectors can be executed in parallel, the circuit depth is governed by the cost of a single sector rather than by the sum over both sectors. Consequently, the initial-state preparation remains polynomial in the number of unit cells.

\subsection{Measurement protocol}

Although the Berry phase is not itself a direct physical observable, it can be inferred from the expectation value of the periodic-boundary-condition-resolved position operator defined in \eqref{X}. 
By applying the Jordan--Wigner transformation in \eqref{JW} together with the qubit-index mapping in \eqref{qubit mapping}, this operator becomes
\begin{equation}\label{X qc}
\hat{X} = \sum_{j=0}^{N-1}  (j+1)\left(n_{2j} + n_{2j+1}+ n_{2N+2j} + n_{2N+2j+1}\right),
\end{equation}
where \(n_j\) denotes the number operator on the \(j\)-th logical qubit. 
Since \(\hat{X}\) is diagonal in the computational basis, all of its components commute with the standard bit-string measurement. 
Therefore, the quantity required for the Berry phase can be estimated from a single set of computational-basis measurements.

Let $\ket{b}\equiv \ket{b_0\dots b_{4N-1}}$ denote a computational-basis bit string, where \(b_j\in\{0,1\}\), and define \(p(b)=|\inner{b}{\psi}|^2\). Using the spectral decomposition in the computational basis, the quantity \(z_N[\tilde{n}/\tilde{N}]\) in \eqref{z} can be written as
\begin{align}
z_N[\tilde{n}/\tilde{N}]
&= \sum_{bb'}\inner{\psi}{b}\!\!\bra{b}e^{\frac{i2\pi \tilde{N}}{N} \hat{X}}\ket{b'}\!\!\inner{b'}{\psi} \notag  \\
&= \sum_b p(b) e^{\frac{i2\pi  \tilde{N}}{N}\bra{b}\hat{X}\ket{b}},
\end{align}
with \(\bra{b} n_j \ket{b} = b_j.\)
The Berry phase is then obtained from the complex phase of \(z_N[\tilde{n}/\tilde{N}]\), as specified in \eqref{gamma}.

In practice, however, it is unnecessary, and generally inefficient for large systems, to store all sampled bit strings and reconstruct the full empirical distribution \(p(b)\). 
Instead, \(z_N[\tilde{n}/\tilde{N}]\) can be estimated directly by averaging the phase factor associated with each measurement outcome. 
Let \(\ket{b^{(m)}}\) denote the \(m\)-th bit string obtained from the quantum device. 
For each measurement outcome, we compute
\begin{equation}
z_N^{(m)}[\tilde{n}/\tilde{N}] =  e^{\frac{i2\pi  \tilde{N}}{N}\bra{b^{(m)}}\hat{X}\ket{b^{(m)}}}.    
\end{equation}
After \(M\) measurements, the estimator of \(z_N[\tilde{n}/\tilde{N}]\) is given by
\begin{equation}
\bar{z}_N[\tilde{n}/\tilde{N}] = \frac{1}{M} \sum_{m=1}^{M} z_N^{(m)}[\tilde{n}/\tilde{N}].
\end{equation}
Accordingly, the Berry phase is estimated as
\begin{equation}
\bar{\gamma} = \im\ln\bar{z}_N[\tilde{n}/\tilde{N}],
\end{equation}
where the result is understood modulo \(2\pi\).

The electron polarization can be estimated from the same computational-basis measurements. For each sampled bit string \(\ket{b^{(m)}}\), the local sublattice-resolved charge imbalance at unit cell \(j\) is evaluated as
\begin{equation}
 p^{\mathrm{e},(m)}_{j} = \bra{b^{(m)}}  n_{2j} + n_{2N+2j} - n_{2j+1} - n_{2N+2j+1} \ket{b^{(m)}},
\end{equation}
for $j = 0,\dots, N-1$. The corresponding estimator is then
\begin{equation}
\bar{p}^{\mathrm{e}}_{j} = \frac{1}{M}\sum_{m=1}^{M}  p^{\mathrm{e},(m)}_{j}.
\end{equation}

Thus, both the Berry phase and the electron polarization can be obtained from standard computational-basis measurements. 
The Berry phase is extracted by classically post-processing the sampled bit strings into the complex estimator \(\bar{z}_N[\tilde{n}/\tilde{N}]\), whereas the electron polarization is obtained by directly averaging the corresponding occupation-number imbalance. 
In contrast to generic observables such as the ground-state energy, whose Pauli-string decomposition generally requires measurements in multiple bases, the present quantities are diagonal in the computational basis after the Jordan--Wigner transformation. 
Consequently, their estimation requires neither additional basis rotations nor separate measurement settings, and therefore introduces no extra circuit depth beyond that required for state preparation and adiabatic evolution.

\section{Numerical simulations}\label{sec:simulation}
We implement and simulate the quantum adiabatic simulation for the one-dimensional SSHH model within the Qiskit framework~\cite{qiskit2024}. 
To resolve the relevant topological signatures within the limits of classical circuit simulation, we use \(N=6\) unit cells, corresponding to \(24\)-qubit circuits. 
The Berry phase is evaluated for the half-filled periodic SSHH ground state with \(12\) electrons, whereas the sublattice polarization is computed for the open-boundary ground state with \(14\) electrons. 
The results show that the SSH topological signatures persist under weak Hubbard interactions but break down under sufficiently strong chiral-symmetry-breaking interactions.

\subsection{Validation of the Trotter decomposition}
Before evaluating the topological observables, we first examine the validation of the first-order Trotter decomposition with respect to the total evolution time \(T\) and the number of propagator intervals \(L\). 
This analysis fixes the simulation setting used in the subsequent Berry-phase and sublattice-polarization calculations.

We assess convergence using two fidelity criteria. 
Without Hubbard interactions, the circuit-simulated final state is compared with the analytically predicted final state. 
With Hubbard interactions, where no analytical reference state is readily available, we instead compare final states obtained from adjacent values of the propagator interval number \(L\).

For both assessments, we fix the hopping amplitudes at \(v=0.5\) and \(w=1.5\). 
The total evolution time is varied over \(T=1,5,15,\) and \(80\), while the number of intervals is chosen from
\[
L \in \{1,10,20,30,40,50,60,80,100,120,150\}.
\]
Note that, for the interacting case, the fidelity obtained at $L=1$ is computed with the initial state, severing as a reference baseline. 

\begin{figure}[t]
\centering
\includegraphics[width=0.8\linewidth]{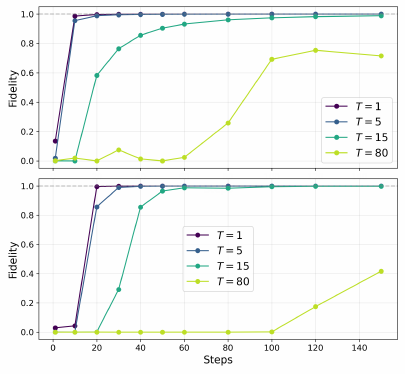}
\caption{Final-state fidelity for the case with \(U=0\) (top) and the case with \(U=1\) (bottom).}
\label{Trotter fidelity}
\end{figure}

The results for the noninteracting case \(U=0\) and the interacting case \(U=1\) are shown in the top and bottom panels of \figref{Trotter fidelity}, respectively. 
For short evolution times, \(T=1\) and \(T=5\), the final-state fidelity rapidly approaches unity, with near-ideal values already achieved at relatively small \(L\). 
For the intermediate evolution time \(T=15\), convergence with respect to \(L\) is slower; nevertheless, the fidelity increases over most of the sampled range and becomes close to unity at \(L=150\). 
In contrast, for the long evolution time \(T=80\), the chosen range of interval numbers is clearly insufficient, as the fidelity remains substantially below unity even at the largest value of \(L\) considered.

These results indicate that longer evolution times require a larger number of intervals to maintain the accuracy of the adiabatic simulation. We also observe that the interacting case appears to approach unity with fewer intervals than the noninteracting case. However, this behavior should not be interpreted as evidence that the interacting simulation is more accurate in an absolute sense. In the interacting case, the fidelity is evaluated between final states obtained at two adjacent entries in the chosen sequence of interval numbers, rather than between the simulated state and an exact reference state. Thus, this quantity only diagnoses the stability of the final state under increasing \(L\); a value close to unity indicates that the result is insensitive to further refinement of \(L\), but does not by itself guarantee proximity to the ideal adiabatically evolved state.

Based on these observations, we choose \(T=1\) and \(L=40\) as the simulation setting for the subsequent numerical experiments. This choice provides high fidelity in the noninteracting benchmark and stable behavior in the interacting case, while keeping the circuit depth within a tractable range.

\subsection{Many-body Berry phase}
Next, we investigate the Berry phase of the SSHH model for different values of \(\Delta U\). 
We fix \(U_A=0.01\) and the intracell hopping amplitude \(v=1\), while varying the intercell hopping amplitude \(w\) from \(0\) to \(2\) in increments of \(0.25\).
In addition, we include two sampling points near the transition, \(w=0.99\) and \(w=1.01\), while omitting the critical point \(w=1\), where the Berry phase is ill-defined because the gap closes.

\begin{figure}[t]
    \centering
    \includegraphics[width=0.8\linewidth]{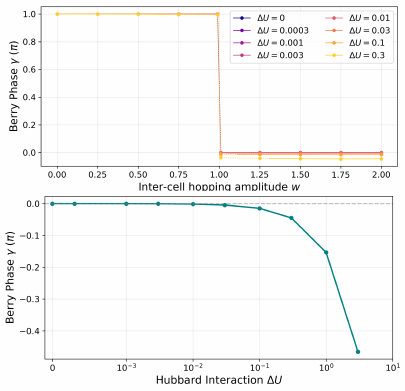}
    \caption{Many-body Berry phase for different values of $\Delta U$.}
    \label{BerryPhase_2in1}
\end{figure}

The results are shown in the top panel of \figref{BerryPhase_2in1}. 
For \(\Delta U=0\), the model reduces to the SSH model, and the Berry phase exhibits a sharp step-like change across \(w=1\). 
As \(\Delta U\) is increased from \(0.0003\) to \(0.3\), this step-like signature is progressively diminished due to the breaking of chiral symmetry. 
Moreover, in the both topologically trivial and nontrivial regime, the Berry phase is no longer pinned to the quantized value \(\gamma=1\) and \(\gamma=0\), respectively, in our phase convention. 
This behavior indicates that the topological phase breaks down when the symmetry protection responsible for Berry-phase quantization is removed.

To further assess the robustness of the Berry phase in the weak \(\Delta U\)  regime, we fix \(w=1.5\) and extend the set of \(\Delta U\) toward smaller values. 
The results are shown in the bottom panel of \figref{BerryPhase_2in1}.
In the present setting, the Berry phase remains essentially unchanged for \(\Delta U \lesssim 10^{-2}\). 
This indicates that the interaction imbalance acts only as a weak perturbation in this regime, so that the topological signature is preserved.
In contrast, for \(\Delta U \gtrsim 10^{-1}\), the Berry phase is shifted appreciably away from the quantized value \(\gamma=0\), indicating that the symmetry-breaking perturbation is no longer negligible.
This trend agrees with the theoretical expectation that Berry-phase quantization is robust only when the unbalanced Hubbard interaction remains sufficiently weak.

\subsection{Spatial profile of sublattice polarization}
We finally examine the spatial profile of the sublattice polarization in the open-boundary SSHH model and its response to \(\Delta U\). 
Following the Berry-phase analysis, we use the same sets of $U_A$ and \(\Delta U\) values, but set \(v=0.1\) and \(w=1.0\) to enhance the dimerization so as to better separate edge and bulk polarization features in 6-unit-cell systems.

\begin{figure}[t]
\centering
\includegraphics[width=0.8\linewidth]{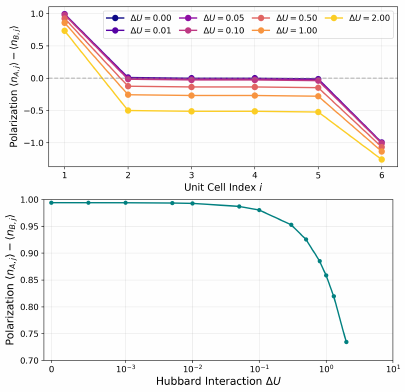}
\caption{Sublattice polarization for different values of $\Delta U$.}
\label{Polarization_2in1}
\end{figure}

As shown in the top panel of \figref{Polarization_2in1}, for \(\Delta U=0\), the spatial profile of the sublattice polarization is localized at the two edges, while the bulk remains essentially unpolarized. 
When \(\Delta U\neq 0\), this profile is progressively distorted, and a finite polarization offset develops in the bulk.

To resolve the small-imbalance regime more clearly, we further examine smaller values of \(\Delta U\), focusing on the response at the first unit cell, namely the edge on the $A$ sublattice.
The results are shown in the bottom panel of \figref{Polarization_2in1}.
For \(\Delta U \lesssim 0.01\), the edge-localized feature is retained, with the corresponding sublattice polarization remaining close to unity.  
For larger values, \(\Delta U \gtrsim 0.5\), the sublattice polarization gradually degrades as $\Delta U$ increase.

These results indicate that the spatial profile of the sublattice polarization retains its edge-localized character as long as the interaction imbalance $\Delta U$ is sufficiently weak.

\section{Summary}

We have investigated the topological properties of the SSHH model in detail, focusing on the many-body Berry phase under PBC and the spatial profile of sublattice polarization under OBC. We also examined the feasibility and limitations of obtaining the true many-body ground state of the SSHH model via quantum adiabatic simulation, starting from an initial state constructed from pre-solved one-particle states of the SSH model. Subtle issues such as spin-population conservation and chiral-symmetry breaking were analyzed and clarified.

Using the Jordan--Wigner transformation, the many-body wavefunction of the SSHH model on a lattice with $N$ unit cells can be mapped to a quantum register of $4N$ qubits. Based on a first-order Trotter decomposition, we constructed explicit quantum circuits implementing the adiabatic time evolution, together with a systematic circuit-based preparation of the required initial states. We also proposed practical measurement protocols for extracting the many-body Berry phase and the spatial profile of sublattice polarization.

We performed classical numerical simulations of the proposed circuits using 24 qubits for a 6-unit-cell SSHH model. Under PBC, the many-body Berry phase retains the expected step-like behavior, pinned to \(0\) and \(\pi\), as long as the imbalanced Hubbard interaction \(\Delta U\) remains weak. As \(\Delta U\) exceeds a threshold, this topological signature begins to break down as the Berry phase deviates from these quantized values. Under OBC, the spatial profile of the sublattice polarization remains edge-localized for weak \(\Delta U\), but this localization gradually degrades once \(\Delta U\) crosses the threshold.

Taken together, these results demonstrate that adiabatic quantum simulation provides a practical route for probing physically meaningful observables in interacting many-body systems beyond ground-state energy estimation. Although the present results are obtained from classical simulations of small systems ($N=6$), the required qubit number ($4N$), gate complexity, measurement shots, and classical pre- and post-processing costs all scale polynomially with system size. Our work therefore establishes a proof-of-concept framework for studying nontrivial topological and spatial properties of interacting many-body systems via adiabatic quantum simulation, pointing toward potential implementations for large-size systems on future large-scale quantum computers.

\section*{AI Tool Usage Statement}
ChatGPT and Gemini were used for text refinement and Python code support. All content, code, and results were reviewed and validated by the authors, who take full responsibility for the final work.
\bibliographystyle{IEEEtran} 
\bibliography{reference}

\end{document}